\documentstyle[12pt]{article}

\begin{document}
{\sf \begin{center} \noindent
{\Large \bf Exponential stretching in filaments as fast dynamos in Euclidean and curved Riemannian 3D spaces}\\[3mm]

by \\[0.3cm]

{\sl L.C. Garcia de Andrade}\\

\vspace{0.5cm} Departamento de F\'{\i}sica
Te\'orica -- IF -- Universidade do Estado do Rio de Janeiro-UERJ\\[-3mm]
Rua S\~ao Francisco Xavier, 524\\[-3mm]
Cep 20550-003, Maracan\~a, Rio de Janeiro, RJ, Brasil\\[-3mm]
Electronic mail address: garcia@dft.if.uerj.br\\[-3mm]
\vspace{2cm} {\bf Abstract}
\end{center}
\paragraph*{}
A new antidynamo theorem for non-stretched twisted magnetic flux
tube dynamo is obtained. Though Riemannian curvature cannot be
neglected since one considers curved magnetic flux tube axis, the
stretch can be neglect since one only considers the limit of thin
magnetic flux tubes. The theorem states that marginal or slow
dynamos along curved (folded), torsioned (twisted) and non-stretched
flux tubes endowed with diffusionless plasma flows, if a constraint
is imposed on the relation between poloidal and toroidal magnetic
fields in the helical dynamo case. A formula for the stretch of flux
tubes is derived. From this formula one shows that the Riemann flux
tube is stretched by an interaction between the plasma flow
vorticity and torsion, in accordance with our physical intuition.
Marginal diffusionless dynamos are shown to exist obtained in the
case of flux tube dynamos exponential stretching. Thus slow dynamos
can be obtaining on the flux tube under stretching. Filamentary
dynamos anti-dynamos are also considered. As flux tubes possess a
magnetic axis torsioned filament, it can also be considered as
thegerm of a fast dynamo in flux tubes Riemannian curved space. It
is shown that for non-stretched filaments only untwist and unfold
filaments can provide dynamo action in diffusive case. A condition
for exponential stretching and fast dynamos in filaments is given.
These results are actually in agreement with Vishik argument that
fast dynamo cannot be obtained in non-stretched flows. Actually the
flux tube result is the converse of Vishik's lemma.{\bf PACS
numbers:\hfill\parbox[t]{13.5cm}{02.40.Hw:differential geometries.
91.25.Cw-dynamo theories.}}

\newpage
\newpage
 \section{Introduction}
 Earlier a theorem stated and proved by Zeldovich (anti-dynamo theorem) \cite{1} showed that planar flows cannot support
 dynamo action. Along with Cowling큦 theorem \cite{2} which has been recently tested for black holes by Brandenburg \cite{3}, these are
 the two main and more traditional and well-tested antidynamo theorems ever. Actually in dynamo action one of the main ingredients
 is exponential stretching , so-well investigated by Friedlander and Vishik \cite{4}. In this same paper they argued that there
 are several topological obstructions to the existence of Anosov flows \cite{5} in Riemannian 3-D space. Anosov flows
 which are Riemannian spaces of constant curvature endowed with geodesic flows, have been shown recently by Chicone
 and Latushkin \cite{6} to be simple fast dynamos in compact Riemannian manifolds. Other types of fast dynamo mechanisms with stretching flux tubes
 in Riemannian conformal manifolds have been obtained by Garcia de Andrade \cite{7}. Yet much earlier M. Vishik \cite{8}
 has argued that only slow dynamos can be obtained from non-stretching dynamo flows and no fast dynamos so
 well-known to be obtained from the  Vainshtein and Zeldovich \cite{9} work on stretch-twist and fold (STF) \cite{10}
 magnetic dynamo generation mechanism could be obtained in this way. This sort of "anti-fast-dynamo" theorem on flows can here
 be generalize to flux tube dynamos and filaments. Actually in this paper we show that the Vishik argument can be extended
 as to provide a anti-fast-dynamo theorem for filaments and tubes in some particular cases and subject to bounds in poloidal and
 toroidal magnetic fields and the twist (torsion) of the magnetic flux tube axis. Exponential stretching of
 tube condition is derived and from it one is able to derive our
 results. Non-Stretching in diffusionless media is used for flux
 tube dynamos while diffusive filaments are used in the second case.
 Marginal dynamos are obtained in the case of steady dynamos, while
 constraints are placed on the magnetic fields for dynamo action to
 be effective.  Diffusion processes have been previously also investigated in the context of Riemannian geometry
 by S. Molchanov \cite{11}. Such slow dynamos which have previously obtained by Soward \cite{12} which also argued that
 fast dynamos actions would be possible in regions where no non-stretching flows would be presented,such in some curved surfaces.
 These surfaces could be of course Riemannian. In the case of filaments it is shown that dynamo action for non-stretched
 filaments would be obtained if they were unfolded and untwisted as well. When torsion vanishes
 filaments are planar and by Zeldovich anti-dynamo theorem \cite{1} cannot support dynamo action, actually it is shown that it
 generates a static magnetic initial field and a steady perturbation which may be a marginal dynamo at maximum.
 Thus we may conclude that in the same way fast dynamo are generated by stretch, folding
 and twisting of the loops or filaments it seems that non-stretched
 ,fold and twisted filaments leads to slow dynamos filaments. The paper is organized as
 follows: In section 2 a brief review on dynamics of holonomic Frenet frame is presented with the discussion of
 vortex-filament exponential stretching. In section
 3 the self-induction equation is solved in the framework of the Ricca큦 \cite{13} twisted magnetic flux tube in Riemannian
 3D manifold and anti-dynamo theorem presented. In section 4 a
 twisted (torsioned) and curved filament is shown to be a fast
 dynamo in Euclidean space. Section 5 contains discussions and
 future prospects.
\newpage
\section{Exponential stretching in dynamo flux tube} This section contains a very brief review
of the Serret-Frenet holonomic frame \cite{14} equations that are
specially useful in the investigation of STF Riemannian flux tubes
in magnetohydrodynamics (MHD) with magnetic diffusion. Here the
Frenet frame is attached along the magnetic flux tube axis which
possesses Frenet torsion and curvature , which completely determine
topologically the filaments, one needs some dynamical relations from
vector analysis and differential geometry of curves such as the
Frenet frame $(\textbf{t},\textbf{n},\textbf{b})$ equations
\begin{equation}
\textbf{t}'=\kappa\textbf{n} \label{1}
\end{equation}
\begin{equation}
\textbf{n}'=-\kappa\textbf{t}+ {\tau}\textbf{b} \label{2}
\end{equation}
\begin{equation}
\textbf{b}'=-{\tau}\textbf{n} \label{3}
\end{equation}
The holonomic dynamical relations from vector analysis and
differential geometry of curves by
$(\textbf{t},\textbf{n},\textbf{b})$ equations in terms of time
\begin{equation}
\dot{\textbf{t}}=[{\kappa}'\textbf{b}-{\kappa}{\tau}\textbf{n}]
\label{4}
\end{equation}
\begin{equation}
\dot{\textbf{n}}={\kappa}\tau\textbf{t} \label{5}
\end{equation}
\begin{equation}
\dot{\textbf{b}}=-{\kappa}' \textbf{t} \label{6}
\end{equation}
along with the flow derivative
\begin{equation}
\dot{\textbf{t}}={\partial}_{t}\textbf{t}+(\vec{v}.{\nabla})\textbf{t}
\label{7}
\end{equation}
From these equations and the generic flow \cite{13}
\begin{equation}
\dot{\textbf{X}}=v_{s}\textbf{t}+v_{n}\textbf{n}+v_{b}\textbf{b}
\label{8}
\end{equation}
one obtains
\begin{equation}
\frac{{\partial}l}{{\partial}t}=(-\kappa{v}_{n}+{v_{s}}')l\label{9}
\end{equation}
where l is given by
\begin{equation}
l:=(\textbf{X}'.\textbf{X}')^{\frac{1}{2}}\label{10}
\end{equation}
\begin{equation}
l=l_{0}e^{\int{(-\kappa{v}_{n}+{v_{s}}')dt}}\label{11}
\end{equation}
which shows that stretching which shows that if $v_{s}$ is constant,
which fulfills the solenoidal incompressible flow
\begin{equation}
{\nabla}.\textbf{v}=0\label{12}
\end{equation}
and $v_{n}$ vanishes, one should have an non-stretched twisted flux
tube. This is exactly the choice $\textbf{v}=v_{0}\textbf{t}$, where
$v_{0}=constant$ is the steady flow one uses here. This definition
of magnetic filaments is shows from the solenoidal carachter of the
magnetic field
\begin{equation}
{\nabla}.\textbf{B}=0\label{13}
\end{equation}
where $B_{s}$ is the toroidal component of the magnetic field. In
the next section one shall solve the diffusion equation in the
steady case in the non-holonomic Frenet frame as
\begin{equation}
{\partial}_{t}\textbf{B}={\nabla}{\times}(\textbf{v}{\times}\textbf{B})+{\eta}{\nabla}^{2}\textbf{B}
\label{14}
\end{equation}
where ${\eta}$ is the magnetic diffusion. Since in astrophysical
scales ,
${\eta}{\nabla}^{2}\approx{{\eta}{L^{-2}}}\approx{{\eta}{\times}10^{-20}}cm^{-2}$
for a solar loop scale length of $10^{10}cm$ \cite{6} one notes that
the diffusion effects are not highly appreciated in astrophysical
dynamos, though they are not neglected here. Let us now consider the
magnetic field definition in terms of the magnetic vector potential
$\textbf{A}$ as
\begin{equation}
\textbf{B}={\nabla}{\times}\textbf{A} \label{15}
\end{equation}
the gradient operator is
\begin{equation}
{\nabla}=\textbf{t}{\partial}_{s}+\frac{1}{r}\textbf{e}_{\theta}+\textbf{e}_{r}{\partial}_{r}\label{16}
\end{equation}
\newpage
Let us now consider the Riemann metric of a flux tube in curvilinear
coordinates $(r,{\theta},s)$. This metric is encoded into the
Riemann line element
\begin{equation}
ds^{2}= dr^{2}+r^{2}d{{\theta}}^{2}+K^{2}ds^{2} \label{17}
\end{equation}
where, accordingly to our hypothesis,
$K^{2}=(1-{\kappa}rcos{\theta})^{2}$ which contributes to the
Riemann curvature.  These can be easily computed with the computer
tensor package as the Riemann curvature components
\begin{equation}
R_{1313}=R_{rsrs}=
-\frac{1}{4K^{2}}[2K^{2}{\partial}_{r}A(r,s)-A^{2}]=-\frac{1}{2}\frac{K^{4}}{r^{2}}=-\frac{1}{2}r^{2}{\kappa}^{4}cos^{2}{\theta}
\label{18}
\end{equation}
\begin{equation}
R_{2323}=R_{{\theta}s{\theta}s}= -\frac{r}{2}A(r,s)=
-{K^{2}}\label{19}
\end{equation}
where $A:={\partial}_{r}K^{2}$. In the case of thin tubes addressed
here, where $K^{2}(r,s)\approx{1}$, these Riemann curvature reduces
to
\begin{equation}
R_{1313}=R_{rsrs}= -\frac{1}{r^{2}} \label{20}
\end{equation}
This shows that Riemann curvature of the tube is particularly strong
when one approaches the tube. In the next section one applies some
of the mathematical machinery derived in this section to the
formulation of a new anti-dynamo theorem. Let us now consider the
generic flow in the case of flux tube in Riemannian space. As in
Ricca's \cite{12} one considers that no radial components of either
flows or magnetic fields are present during the computations, thus
\begin{equation}
\dot{\textbf{X}}=v_{s}\textbf{t}+v_{\theta}\textbf{e}_ {\theta}
\label{21}
\end{equation}
By considering the relation between the two bases
$(\textbf{e}_{r},\textbf{e}_{\theta},\textbf{t})$ and Frenet frame
as
\begin{equation}
\textbf{e}_{r}=cos{\theta}\textbf{n}+sin{\theta}\textbf{b}\label{22}
\end{equation}
and
\begin{equation}
\textbf{e}_{\theta}=-sin{\theta}\textbf{n}+cos{\theta}\textbf{b}\label{23}
\end{equation}
one may express formula (\ref{21}) as
\begin{equation}
\dot{\textbf{X}}=v_{s}\textbf{t}-v_{\theta}sin{\theta}\textbf{n}+v_{\theta}cos{\theta}\textbf{b}
\label{24}
\end{equation}
Comparison between (\ref{24}) and (\ref{21}) yields
\begin{equation}
v_{n}=-v_{\theta}sin{\theta} \label{25}
\end{equation}
and
\begin{equation}
v_{b}=v_{\theta}cos{\theta} \label{26}
\end{equation}
which by a simple comparison with expression (\ref{22}) yields
\begin{equation}
\frac{{\partial}l}{{\partial}t}=(\kappa{v}_{\theta}sin{\theta}
+{v_{s}}')l\label{27}
\end{equation}
Thus one finally finds out an expression between the exponential
stretching l as
\begin{equation}
l=l_{0}e^{\int{(\kappa{v}_{\theta}sin{\theta}
+{v_{s}}')dt}}\label{28}
\end{equation}
From the solenoidal incompressible flow equation
\begin{equation}
{\nabla}.\textbf{v}=0\label{29}
\end{equation}
one obtains
\begin{equation}
{\partial}_{s}{v}_{\theta}=r{\kappa}{\tau}sin{\theta}{v_{\theta}}\label{30}
\end{equation}
where one has taken the
operator${\partial}_{\theta}=-{\tau}^{-1}{\partial}_{s}$ and use
flux tube definition twist angle
${\theta}={\theta}_{0}-{\int{{\tau}ds}}$. Note that substitution of
this result into expression (\ref{28}) and taking into account that
in Ricca's tube ${v'}_{s}$ vanishes, upon integration leads to
\begin{equation}
l=l_{0}e^{({{{\tau}^{2}}_{0}r\int{sin{2{\theta}}dt})}}\label{31}
\end{equation}
where one has consider to simplify computations that the dynamos are
helical which means that the torsion and Frenet curvature are taken
as equal and constants. By the mathematical operator
${d{t}}=-{{\omega}_{0}}^{-1}d{\theta}$ and substitution into the
formula (\ref{31}) yields upon integration that
\begin{equation}
l=l_{0}e^{({-\frac{{{\tau}^{2}}_{0}r}{2{\omega}_{0}}\int{sin{{\theta}}d{\theta}})}}\label{32}
\end{equation}
and
\begin{equation}
l=l_{0}e^{({\frac{{{\tau}^{2}}_{0}r}{2{\omega}_{0}}{cos{\theta}}})}\label{33}
\end{equation}
where one has taken the following expression into account
\begin{equation}
{\omega}_{0}r=v_{\theta}\label{34}
\end{equation}
where ${\omega}_{0}$ is the constant rotation of the plasma. This
shows that there is a stretching in the tube if $({\omega}_{0}>0)$
and a compression or non-stretching if $({\omega}_{0}<0)$. The
plasma rotation inside the flux tube can be obtained from the
vorticity equation
\begin{equation}
{\nabla}.\vec{{{\omega}}}={\nabla}.{\nabla}{\times}\textbf{v}=0\label{35}
\end{equation}

\newpage

which yields the following PDEs
\begin{equation}
{\omega}_{r}=-{\partial}_{s}v_{\theta}\label{36}
\end{equation}
\begin{equation}
{\omega}_{\theta}={\omega}_{0}=-{\partial}_{r}v_{s}\label{37}
\end{equation}
\begin{equation}
{\omega}_{s}=-[{\partial}_{r}v_{\theta}-\frac{cos{\theta}}{r}{\tau}_{0}v_{\theta}]\label{38}
\end{equation}
From vorticity expression (\ref{37}) one obtains
\begin{equation}
{\omega}_{0}r=-v_{s}\label{39}
\end{equation}
Substitution of the expression for ${\partial}_{s}v_{\theta}$ above
into expression (\ref{36}) now yields
\begin{equation}
{\omega}_{r}=-{{\tau}_{0}}^{2}rsin{\theta}v_{\theta}\label{40}
\end{equation}
for the thin flux tube. For steady dynamos one obtains \cite{18}
\begin{equation}
(\textbf{v}.{\nabla})\textbf{B}=(\textbf{B}.{\nabla})\textbf{v}\label{41}
\end{equation}

\begin{equation}
\frac{B_{\theta}}{B_{s}}=\frac{v_{\theta}}{v_{s}}\label{42}
\end{equation}
By making use of the vorticity equations one obtains
\begin{equation}
\frac{B_{\theta}}{B_{s}}\approx{\frac{1}{{\tau}_{0}rcos{\theta}}}\label{43}
\end{equation}
Note then that near to the flux tube axis $(r\approx{0})$ one
obtains that the toroidal magnetic field decreases from toroidal
field. A dynamo test can be obtained by computing the integral of
Zeldovich
\begin{equation}
\frac{4{\pi}d{\epsilon}_{M}}{dt}=\int{\textbf{B}.(\textbf{B}.{\nabla})\textbf{v}dV}
\label{44}
\end{equation}
where another term of diffusion has been dropped since here
diffusion vanishes. After some algebra one obtains for the flux tube
the following equation
\begin{equation}
\frac{4{\pi}d{\epsilon}_{M}}{dt}=\int{[{B}_{\theta}{{\tau}_{0}}^{2}sin{\theta}[v_{\theta}-{{\tau}_{0}}^{-1}v_{s}]+
B_{s}v_{s}](B_{s}-\frac{B_{\theta}{{\tau}_{0}}^{-1}}{r})dV}
\label{45}
\end{equation}
which showa that
\begin{equation}
B_{s}=\frac{B_{\theta}{{\tau}_{0}}^{-1}}{r} \label{46}
\end{equation}
implies the existence of a marginal dynamo, where $
\frac{4{\pi}d{\epsilon}_{M}}{dt}$ vanishes. Note that this result
does not depend directly on the stretching but on the torsion which
indirectly is responsible for stretching, therefore exponential
stretching may exists even for marginal dynamos, which would
represent a converse result of a Vishik's antidynamo theorem for
flux tubes. Non-stretching flows implies necessarily slow dynamos,
but slow dynamos do not imply necessarily non-stretching. The weak
torsion approximation used here,is in agreement with the torsion
(twist) in the solar twisted coronal loop torsion
$({\tau}_{0}\approx{10^{-10}cm^{-1}})$ \cite{15}.
\newpage

\section{Anti-dynamo theorem in non-stretching filaments}
In this section one considers the anti-dynamo formulation of
non-stretching dynamo flows in twisted filaments. This can be done
by simply considering the gradient along the filament as
${\nabla}=\textbf{t}{\partial}_{s}$ and computing the total magnetic
energy on a diffusive medium as
\begin{equation}
\frac{4{\pi}d{\epsilon}_{M}}{dt}=\int{{B_{s}}^{2}v_{s}\textbf{t}.\textbf{n}dV}:=0
\label{47}
\end{equation}
since $\textbf{n}.\textbf{t}=0$. Here one has considered that
$\textbf{B}:=B_{s}\textbf{t}$. Thus one must say that in a diffusive
media since ${v'}_{s}=0$ and $v_{n}$ also vanishes by assumption
that the following lemma has been proved.

\textbf{lemma}:

Non-stretching vortex filaments in a diffusionless media gives rise
to a marginal or slow dynamo. No fast dynamo being possible.

This can be considered as a sort of anti-fast dynamo theorem from
Vishik's idea. Now we shall relax the idea of diffusionless media
and introduce diffusion into the problem. As note previously by
Zeldovich \cite{1} this leads us to the dynamo action and possibly
to fast dynamos. In the case of diffusive filaments one notes that
the magnetic induction equation
\begin{equation}
\frac{d}{dt}\textbf{B}=(\textbf{B}.{\nabla})\textbf{B}+{\eta}{\nabla}^{2}\textbf{B}
\label{48}
\end{equation}
where ${\eta}$ is the magnetic resistivity or diffusion, which here
one considers as constant, reduces to three scalar equations
\begin{equation} \frac{d{B}_{s}}{dt}=-{\eta}{\kappa}{B}_{s} \label{49}
\end{equation}
\begin{equation} {\kappa}'={\eta}{\kappa}\tau \label{50}
\end{equation}
\begin{equation} -\kappa{\tau}={\eta}{\kappa}'+v_{s} \label{51}
\end{equation}
Solution of these equations can be easily obtained as
\begin{equation} {B}_{s}=exp[-{\eta}\int{{\kappa}^{2}ds}]{B}_{0} \label{52}
\end{equation}
where $\int{{\kappa}^{2}ds}$ is the total Frenet curvature energy
integral. The remaining solutions are
\begin{equation} {\kappa}=exp[{\eta}\int{{\tau}ds}]{\kappa}_{0} \label{53}
\end{equation}
where $\int{{\tau}ds}$ is the total torsion, and in the case of
helical filaments where torsion equals curvature and are constants,
\begin{equation} -{{\tau}_{0}}^{2}=v_{s} \label{54}
\end{equation}
Note that the decaying or not of the magnetic field depends on the
sign of the integral but since the average value of this integral is
positive the magnetic field decays and no dynamo action is possible,
even slow dynamos. Thus in the case of non-stretching filaments, the
presence of diffusion actually enhances the non-dynamo carachter in
Vishik's lemma.
\section{Filamentary fast dynamos in Euclidean space}
Earlier Arnold has argued \cite{16} that no fast dynamo are usually
found in Euclidean 3D spaces. Actually Arnold himself found a
stretching and compressed fast dynamo in curved Riemannian space.
Though this served as motivation for the study of flux tube dynamos
in 3D curved Riemannian space, in this section one shall address the
problem of finding a filamentary fast dynamo in Euclidean space in
3D. The stretching followed by squeezing is a path to finding a
growing magnetic field. Recently Nu\~{n}ez \cite{17} has considered
a similar problem , also making use of Frenet frame as here, by
investigating eigenvalues in plasma flows. In this section one shows
that the stretching condition in filaments is fundamentally
connected to the incompressibilty of the flow. This is simply
understood if one considers the exponent in exponential stretching
in expression
\begin{equation}
{\gamma}:=-\kappa{v}_{n}+{v_{s}}'\label{55}
\end{equation}
which shows that stretching factor gamma is a fundamental quantity
to be examined when one wants to find out a fast dynamo action. Note
that when ${\gamma}\ge{0}$ or ${\gamma}<0$, one would have
respectively either a fast or slow or marginal dynamo and a decaying
magnetic field as found before. Let us now drop the constraint that
$v_{s}$ and substitute the flow
\begin{equation}
\textbf{v}={v}_{n}\textbf{n}+{v_{s}}\textbf{t}\label{56}
\end{equation}
into the solenoidal incompressible flow
\begin{equation}
{\nabla}.\textbf{v}=0\label{58}
\end{equation}
and $v_{n}$ vanishes, one should have an non-stretched twisted flux
tube. This is exactly the choice $\textbf{v}=v_{0}\textbf{t}$, where
$v_{0}=constant$ is the steady flow one uses here. This definition
of magnetic filaments is shows from the solenoidal carachter of the
magnetic field
\begin{equation}
{\nabla}.\textbf{v}=-\kappa{v}_{n}+{v_{s}}'=0\label{57}
\end{equation}
This result is exactly ${\gamma}=0$ which implies no stretching at
all! This lead us to note that if a fast filamentary dynamo action
be possible, a modification of the flow has to be performed. To
investigate this possibility one considers the following form of the
dynamo flow
\begin{equation}
\textbf{v}={v}_{n}\textbf{n}+{v_{s}}\textbf{t}+v_{0}\textbf{b}\label{58}
\end{equation}
which minimally generalizes (\ref{56}). In this case expression for
${\gamma}$ does not vanish and is equal to
\begin{equation}
{\gamma}=-\kappa{v}_{n}+{v_{s}}'={\tau}_{0}v_{0}\label{59}
\end{equation}
Note that again for ${\gamma}>0$ one obtains a fast dynamo since
${\eta}=0$ already and stretching is possible if ${\tau}_{0}$ and
$v_{0}$ possesses the same sign. Actually this flow leads to the
following three scalar dynamo equations
\begin{equation}
\frac{d}{dt}{B}_{s}={\gamma}{B}_{s}-{{\tau}_{0}}^{2}B_{n} \label{60}
\end{equation}
\begin{equation}
\frac{d}{dt}{B}_{n}={\tau}_{0}(v_{s}-{\tau}_{0}-\frac{1}{{\tau}_{0}}{\partial}_{s}v_{n}){B}_{s}
\label{61}
\end{equation}
\begin{equation}
{B}_{s}{{\tau}_{0}}v_{n}=0 \label{62}
\end{equation}
Here ${\gamma}=({v_{s}}'-{\tau}_{0}v_{n})$ since we are cosidering
helical dynamo filaments. From equation (\ref{62}) one obtains
$v_{n}=0$ which simplifies much the other equations. Since in
astrophysical scales , torsion is very weak as happens in
${{\tau}_{0}}^{2}\approx{{\eta}{\times}10^{-20}}cm^{-2}$ for a solar
coronal loop scale, the terms proportional to torsion squared may be
dropped. In this approximation a fast dynamo solution is found as
\begin{equation}
{B}_{s}=B_{0}e^{{\gamma}t} \label{63}
\end{equation}
and
\begin{equation}
\frac{d}{dt}{B}_{n}={\tau}_{0}(v_{s}-{\tau}_{0}){B}_{s} \label{64}
\end{equation}
which yields
\begin{equation}
{B}_{n}=\frac{1}{{v}_{0}}(v_{s}-{\tau}_{0})e^{{\gamma}t} \label{65}
\end{equation}
where to obtain $B_{n}(t,s)$ one made use of the stationary property
of the flow ${\partial}_{t}v_{s}=0$, in order to being able to
integrate (\ref{64}). Thus a fast dynamo action for filamentary
flows in Euclidean 3D space was obtained. Note that this action is
natural and somewhat expected since when we change the dynamo flow
we obtain a three-dimensional $(v_{s},v_{n},v_{b})$ from a two
dimensional flow $(v_{s},v_{n})$, which is strictly forbbiden from
Zeldovich anti-dynamo theorem. Besides if one integrates the flow
equation (\ref{59}) yields
\begin{equation}
{v_{s}}={\tau}_{0}v_{0}s+c_{1}\label{66}
\end{equation}
where $c_{1}$ is an integration constant. Expression (\ref{66})
indicates that the flow along the filament undergoes a uniform
strain \cite{18} due to the stretching of the magnetic filament. Now
if one uses these relations into the total magnetic energy integral,
yields
\begin{equation}
{\epsilon}_{M}=
\frac{a^{2}}{8}[{B_{0}}^{2}s+\frac{{\tau}_{0}}{3}(v_{s}-v_{0})^{3}v_{0}]e^{{\tau}_{0}v_{0}t}\label{67}
\end{equation}
where one has put $c_{1}:=0$  for convenience. This magnetic energy
indicates that the stretching of the magnetic field accumulates
energy and gives rise to a fast dynamo as long as $v_{s}\ge{v_{0}}$
is bounded from below. Physically this means that the flow along the
magnetic filament is stronger than any magnetic orthonormal
perturbation and the system might be stable.
\section{Conclusions} Vishik's idea that the non-stretched dynamos
cannot be fast is tested once more here, by showing that the
fluctuations modes of the solar dynamos are strongly suppressed when
the long wavelength dynamo modes, or small dynamo wave numbers are
effective. Stretching are therefore fundamental in the effective
dynamo convective region of the Sun \cite{6}. The twist or Frenet
torsion is also small, and slow dynamo are shown to be present in
this astrophysical loops. By following the analogy proposed by
Friedlander and Vishik in dynamo theory between vorticity equations
and dynamo equations and considering exponential stretching one
shows that flux tube dynamos are in complete agreement with Ricca's
original Riemannian magnetic flux flux tube model. STF
Zeldovich-Vainshtein  fast dynamo generation method ,is not the only
Riemannian method that can be applied as in Arnold's cat map but
other conformal fast kinematic dynamo models as the conformal
Riemannian one has been recently obtained. Small scale dynamos in
Riemannian spaces can therefore be very useful for our understanding
of more large scale astrophysical dynamos. Other applications of
plasma filaments such as stretch-twist and fold fractal dynamo
mechanism which are approximated Riemannian metrics have been
recently put forward by Vainshtein et al \cite{14}. Finally one has
shown that the Vishik's result is strongly enhanced and no fast
dynamo or even slow dynamo is found and decay of the magnetic field
in non-stretched filaments in presence of magnetic diffusion and the
Frenet torsion is found. As considered by Arnold \cite{16} and
Zeldovich et al \cite{18} no fast dynamos exists in 3D Euclidean
space and a fast dynamo was obtained in 3D curved Riemannian space
by Arnold \cite{16} where stretching in some directions is
compensated by compression in other. In this paper a fast
filamentary dynamo was obtained in Euclidean 3D space. Note that
when torsion vanishes in the last section a marginal dynamo is
obtained, this result, a sort of filamentary antidynamo theorem ,
has been recently obtained \cite{19} in the non-holonomic frame.
\section{Acknowledgements}
I am deeply greatful to Renzo Ricca and Yuri Latushkin for their
extremely kind attention to our work. Thanks are also due to I thank
financial supports from Universidade do Estado do Rio de Janeiro
(UERJ) and CNPq (Brazilian Ministry of Science and Technology).

  \end{document}